\documentstyle[aps,multicol,prl,epsf]{revtex}
\begin{document}
\draft
\title{Persistent currents in  interacting Aharonov-Bohm
interferometers\\ and their enhancement by acoustic radiation}
\author{O. Entin-Wohlman$^{a,c}$,  Y. Imry$^b$, and A.
Aharony$^a$}
\address{$^a$School of Physics and Astronomy, Raymond and Beverly Sackler
Faculty of Exact Sciences, \\ Tel Aviv University, Tel Aviv 69978,
Israel\\ }
\address{$^b$Department of Condensed Matter Physics, The Weizmann
Institute of Science, Rehovot 76100, Israel}
\address{$^c$Albert Einstein
Minerva Center for Theoretical Physics at the Weizmann Institute
of Science, Rehovot 76100, Israel}
\date{\today}
\maketitle
\begin{abstract}
We consider an Aharonov-Bohm interferometer,  connected to two
electronic reservoirs, with a quantum dot  embedded on one of its
arms. We find a general expression for the persistent current at
steady state, valid for the case where the electronic system is
free of interactions except on the dot. The result is used to
derive the modification in the persistent current brought about by
coupling the quantum dot to a phonon source. The magnitude of the
persistent current is found to be enhanced  in an appropriate
range of the intensity of the acoustic source.

\end{abstract}

\pacs{PACS numbers: 73.23.-b, 72.15.Gb, 71.38.-k}

\begin{multicols}{2}

It has  long been established, both theoretically \cite{buttiker}
and experimentally \cite{levy}, that a magnetic flux which threads
an isolated electronic system creates a persistent current at
thermal equilibrium. That current, which is equivalent to the
thermodynamic orbital magnetic moment of the electrons, arises
from the interference of the electronic wave functions, and, as
long as the electrons are phase-coherent, survives the presence of
reasonably strong static disorder \cite{joe}. When the electrons
are coupled to a phonon bath, the naive expectation is that the
persistent current will diminish, due to loss of coherence, caused
by inelastic processes as well as by renormalization effects due
to the ``dressing" of the electrons by the phonons. However, it
turns out that this is not the whole effect. For example, in the
case of strongly localized electrons, the coupling to the acoustic
waves gives rise to an additional persistent current
\cite{aronov}, brought about by delicate resonant processes
\cite{holstein}, which result in a non-monotonic temperature
dependence of the orbital magnetic moment at sufficiently low
temperatures. Somewhat related examples are the enhancement of the
persistent current in response to an external ac electric noise
\cite{kravtsov,mohanty}, and its peculiar behavior under the
effect of a nonequilibrium electron energy distribution
\cite{chalaev} or of a periodic oscillating potential
\cite{moskalets}.

Here we consider the persistent current $I_c$, circulating in the
ring of an Aharonov-Bohm interferometer (ABI), connecting two
electronic reservoirs having equal or slightly different chemical
potentials, $\mu_\ell$ and $\mu_r$. This small voltage
$eV=\mu_\ell-\mu_r$ is not the main source for the deviation from
equilibrium in our case. The current is studied also when the
system is strongly out of equilibrium due to a coupling to an
acoustic source, as in \cite{spritz}. In order to retain the
coherence of the conduction electrons, our results apply only for
size-scales small enough to stay coherent at the given
temperature. Moreover, the strength of the acoustic source is
taken to be such that the additional decoherence due to it is not
essential \cite{mohanty}. We take the electronic system to be
non-interacting, except at a single `site' on the ABI, dubbed
`quantum dot', where the electrons can couple to an external
source of sonic (or electromagnetic) waves, and/or experience
electronic interactions. (For simplicity, we assume that this site
has only a single relevant on-site energy level, $\epsilon_d$.)
Under these conditions, we obtain a general expression for the
persistent current circulating around the Aharonov-Bohm ring [Eq.
(\ref{IC}) below]. This expression does not necessitate a
near-equilibrium situation. We then use the result to show that a
phonon source, interacting with the electrons on the quantum dot,
may lead to a considerable enhancement of the persistent current,
and of the related orbital magnetic moment. This
acousto-persistent current exists even when $V=0$. However, the
coupling of the ring to the leads is crucial. Such a relation
between the acoustic wave intensity and the orbital magnetic
moment opens interesting possibilities for future nano-devices.

\vspace{-0.5cm}

\begin{figure}
\leavevmode \epsfclipon \epsfxsize=6.truecm
\vbox{\epsfbox{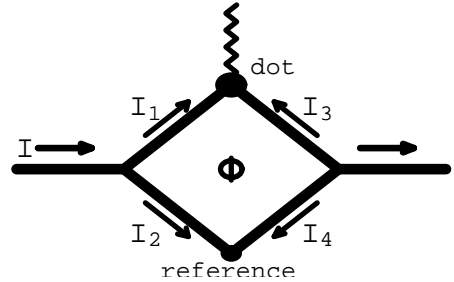}} \vspace{.5cm} \caption{An ABI,
containing a quantum dot on its upper arm and threaded by a
magnetic flux $\Phi$. The wavy line symbolizes the phonons
radiated by the phonon source into the dot. The lower arm of the
ABI contains a `reference' site. The ABI is connected to two
electronic reservoirs whose chemical potentials are either equal
or have a small difference, allowing a current $I$ to flow from
the left to the right.} \label{fig1}
\end{figure}
%\vspace{0.5cm}

Our model system is depicted in Fig. \ref{fig1}. In the steady
state, a dc current, $I$, is flowing through the ABI. To eliminate
spurious circulating currents caused by a geometrical asymmetry of
the ring, we calculate the current $I_{c}$, {\it circulating}
around the ABI ring under the effect of the  Aharonov-Bohm flux
$\Phi$, by
\begin{eqnarray}
I_{c}&=&\frac{1}{2}(I_{1}-I_{2})\Bigl |_{\Phi}-
\frac{1}{2}(I_{1}-I_{2})\Bigl |_{-\Phi}.
\end{eqnarray}
The persistent current with a coupling to one or more electron
reservoirs, but without the additional interactions on the dot,
has been studied in Refs. \cite{butt,claudio,akkermans}.

By using the
%nonequilibrium \cite{comment0}
Keldysh technique
\cite{langreth}, one can find the partial currents flowing in each
arm of the ABI. Moreover, when the electrons are non-interacting,
except on the quantum dot, one is able to express all the relevant
currents in terms of the {\it exact} Green function on the dot,
$G_{dd}$, which includes the interactions and the effect of the
couplings between the dot and the other parts of the circuit. This
Green function has a self-energy which comes from two sources. The
first arises solely from the coupling of the dot to the ring, and
is denoted by $\Sigma_{\rm ext}$. This part can be obtained
straightforwardly, as it pertains to non-interacting electrons.
$\Im \Sigma_{\rm ext}$ represents the lifetime of the on-site
energy level on the dot, turning it into a resonance. The other
part of the self-energy, denoted $\Sigma_{\rm int}$, comes from
the interactions experienced by electrons residing on the dot
\cite{ng}, e.g., the coupling to the phonon source. This
self-energy also depends on the coupling to the other parts of the
circuit. A detailed calculation of $I_{1}$ and $I_{2}$ then yields
\cite{big}
\begin{eqnarray}
I_{c}=\int\frac{d\omega}{i\pi}\frac{f_{\ell}(\omega )+f_{r}(\omega
)}{2}\Bigl [\frac{\partial\Sigma^{A}_{\rm
ext}}{\partial\Phi}G^{A}_{dd}(\omega )-cc\Bigr ],\label{IC}
\end{eqnarray}
where $f_{\ell,r}(\omega )=f(\omega-\mu_{\ell,r})$, with $f(z)
\equiv (e^{\beta z}+1)^{-1}$,
%, $f_{\ell} (f_{r})$
is the electron distribution on the left (right) reservoir, having
the chemical potential $\mu_{\ell}$ ($\mu_{r}$). (It is assumed
that the two reservoirs are otherwise identical. We also use a
unit electron charge, $e=1$, and $\hbar=1$.) The superscripts $A$
($R$) refer to the advanced (retarded) Green functions. Since both
$\Sigma^{A}_{\rm ext}$ and $G^{A}_{dd}$ are even in  $\Phi$, due
to additive contributions (with equal amplitudes) from clockwise
and counterclockwise motions of the electron around the ring (see
e.g. Refs. \onlinecite{claudio,interferometer,amnon}), $I_{c}$ is
odd in $\Phi$, as it should. Another important point to notice is
that the frequency integration in Eq. (\ref{IC}) is restricted to
the band-width of the conduction electrons on the leads. This is
because $G^{A}_{dd}$ attains an imaginary part only due to the
coupling of the dot with the band states on the leads, i.e. for
$\omega$ inside the band.

Several physical remarks on Eq. (\ref{IC}) are called for: (i) The
small voltage allowed on the system is not essential; the result
also holds when $\mu_\ell=\mu_r$. (ii) Equation (\ref{IC})
averages the flux-derivative of the external self-energy over
energy, with weights containing the densities of electrons and
single-particle states with that energy.
% (see Eq. \ref{I0} below).
However, the flux-derivative of the latter does {\em not} appear.
This is reminiscent of the equilibrium case, where the persistent
current is given by the flux-derivative of the energies, weighed
by the populations, without the appearance of the flux-derivative
of those, as in Eq. (\ref{I0}) below (see, for example, Ref.
\cite{aronov}).

Equation (\ref{IC}) holds under four conditions: (i) The
interferometer is {\it symmetric}, i.e., the left and right
couplings of the dot to the interferometer are identical, and so
are those of the reference site (see Fig. \ref{fig1}). Otherwise,
there appears a term proportional to the asymmetry of the ring,
which we omit for clarity; (ii) In order to eliminate the Keldysh
Green function $G^{<}_{dd}$, we have used current conservation,
$I_{1}+I_{3}=0$, in conjunction with the `wide-band' approximation
\cite{jauho}, which implies that $\Sigma_{\rm ext}$ is
energy-independent. (iii)  The chemical potential difference
$\mu_{\ell}-\mu_{r}$ is small, and the system is in the {\it
linear transport regime}. (iv) Last, but not least, is the
requirement that the additional dephasing due to the
nonequilibrium situation can be neglected.

For non-interacting electrons, our expression for $I_{c}$ is
consistent with the result in Ref. \onlinecite{claudio}. It
generalizes the results of Refs. \onlinecite{butt} and
\onlinecite{akkermans} to a steady-state situation. In that
non-interacting case, $\Sigma^A_{\rm ext}$ constitutes the entire
self-energy of the dot Green function, $G^{0A}_{dd}(\omega
)=1/(\omega -\epsilon_{d}-\Sigma^{A}_{\rm ext})$. The persistent
current then becomes
\begin{eqnarray}
I_{c}^{(0)}=\int \frac{2d\omega}{\pi} \frac{f_{\ell}(\omega
)+f_{r}(\omega )}{2} \frac{\partial\delta^{0}(\omega
)}{\partial\Phi},\label{I0}
\end{eqnarray}
where $\delta^{0}$ is the phase of the retarded Green function
$G^R=(G_A)^\ast$ for the non-interacting situation,
\begin{eqnarray}
\tan\delta^{0}(\omega )=-\frac{\Im\Sigma^{R}_{\rm ext}}{\omega
-\epsilon_{d}-\Re\Sigma^{R}_{\rm ext}}.\label{tan}
\end{eqnarray}
This phase is identical to the transmission phase (the Friedel
phase) of the interferometer \cite{ng}. However, this simple
relation between the transmission phase and the persistent current
ceases to hold when the dot self-energy also contains the
interaction-induced part, $\Sigma_{\rm int}$, which  depends on
the flux as well. Thus, {\rm interactions eliminate the simple
relation between the persistent current and the scattering
solution}.

Equation (\ref{IC}), which is our first main result, applies to
{\it any} kind of interactions on the dot, including
electron-electron interactions. In the rest of this Letter we
apply this general result to study the modification of the
persistent current when the dot is coupled to a {\it phonon
source}. Utilizing again the wide-band approximation
\cite{jauho,glazman}, and assuming an interaction between the
phonons and the electron residing on the dot which is {\it linear}
in the phonon coordinates, we find
\begin{eqnarray}
G_{dd}^{R}(\omega )&=&-iK\Bigl
[(1-n_{d})\int_{0}^{\infty}dte^{i(\omega
-\epsilon_{d}-\Sigma^{R}_{\rm
ext})t}e^{\Psi (t)}\nonumber\\
&+&n_{d}\int_{0}^{\infty}dte^{i(\omega
-\epsilon_{d}-\Sigma^{R}_{\rm ext})t}e^{\Psi (-t)}\Bigr ].
\label{Gdd}
\end{eqnarray}
Here $n_{d}$ denotes the electron occupation on the dot
\cite{comment2}. The on-site energy on the dot, $\epsilon_{d}$, is
now renormalized by the polaron shift, $\epsilon_{P}=\sum_{\bf
q}|\alpha_{\bf q}|^{2}/\omega_{q}$, where $\alpha_{\bf q}$ is the
electron-phonon coupling and $\omega_{q}$ denotes the phonon
frequency. Since this renormalization is temperature- and
flux-independent, we will omit it in the following. The other
phonon variables are contained in $K$, the Debye-Waller factor,
and in $\Psi (t)$. Explicitly,
\begin{eqnarray}
K&=&\exp[-\sum_{\bf q}\frac{|\alpha_{\bf
q}|^{2}}{\omega_{q}^{2}}(1+2N_{q})],\nonumber\\
\Psi (t)&=&\sum_{\bf q}\frac{|\alpha_{\bf q}|^{2}}{\omega_{q}^{2}}
[N_{q}e^{i\omega_{q}t}+(1+N_{q})e^{-i\omega_{q}t}],
\end{eqnarray}
where  $N_{q}$ is the phonon occupation of the $q$-mode, which is
not necessarily the thermal equilibrium one, but may be tuned
externally.

Let us now specify to the case of a weak electron-phonon
interaction and not too large $N_{q}$. Expanding Eq. (\ref{Gdd})
to lowest order in $|\alpha_{\bf q}|^{2}$,
%\cite{comment3},
one obtains $G^R_{dd}(\omega )$ in terms of the dot Green function
of the non-interacting case (i.e., in the absence of the
electron-phonon interaction), $G^{0R}_{dd}(\omega )$, at the same
frequency $\omega$, and at frequencies shifted by the phonon
frequencies, $\omega \pm \omega_q$. As a result, the expression
for $I_{c}$ involves the phase $\delta^{0}(\omega )$ at those
frequencies as well. The result is conveniently written in the
form
\begin{eqnarray}
I_{c}=I_{c}^{(0)}+ \Delta I_{c},\label{If}
\end{eqnarray}
where $I_{c}^{(0)}$ is the persistent current of the
non-interacting interferometer, Eq. (\ref{I0}) above, and $\Delta
I_{c}$ is the acousto-persistent current, given, within our
approximation,  by
\begin{eqnarray}
&&\Delta I_{c}=\int \frac{d\omega}{\pi} \frac{f_{\ell}(\omega
)+f_{r}(\omega )}{2}\nonumber\\
&&\times\sum_{\bf q} \Bigl [ A^{+}_{q}\frac{\partial
}{\partial\Phi}\Bigl (\delta^{0}(\omega
+\omega_{q})+\delta^{0}(\omega -\omega_{q})
-2\delta^{0}(\omega )\Bigr )\nonumber\\
&+&A^{-}_{q}\frac{\partial }{\partial\Phi}\Bigl (\delta^{0}
(\omega +\omega_{q})-\delta^{0}(\omega -\omega_{q})\Bigr )\Bigr
 ], \label{ICPH}
\end{eqnarray}
where
\begin{eqnarray}
A_{\bf q}^{+}&=&\frac{|\alpha_{\bf
q}|^{2}}{\omega_{q}^{2}}(1+2N_{q}),\ \ A_{\bf
q}^{-}=\frac{|\alpha_{\bf q}|^{2}}{\omega_{q}^{2}}(2n_{d}-1).
 \label{A}
\end{eqnarray}
The acousto-induced persistent current, $\Delta I_{c}$, consists
of two parts: The first term in Eq. (\ref{ICPH}) is dominated by
the phonon occupations [see Eq. (\ref{A})], via $A_{\bf q}^{+}$.
The second term depends only on the dot's occupation, $n_d$, and
its sign may change according to the relative location of
$\epsilon_d$ with respect to the Fermi energy. The first term in
$\Delta I_{c}$ shows that by shining a beam of phonons of a
specific frequency, the magnitude of the persistent current, and
hence of the orbital magnetic moment of the ABI, can be enhanced
and controlled experimentally, as long as the temperature of the
electronic system and the intensity of the phonon source $N_q$ are
low enough to retain coherent motion of the electrons. This
intensity
%$N_q$
is also limited by our weak effective interaction approximation.
%%%%%%%%%%%%%%%%Our new term yields a current whose maximal order of magnitude for
%our simple model is $(e/h) N^{I}_{ph}\Gamma$. Here $\Gamma$ is the
%width of the dot resonance, and $N^{I}_{ph}$ is the effective
%number of resonant phonons  per mode, weighed by the interaction,
%$ N^{I}_{ph} = \frac{|\alpha_{\bf
%q}|^{2}}{\omega_{q}^{2}}(1+2N_{q})$.
Similar considerations apply to photons. Both the precise
magnitude of these effects and the above bounds depend on the
detailed geometry of the dot and on the acoustic (or
electromagnetic) mismatch. Such calculations go beyond the scope
of the present paper.

Equation (\ref{ICPH}) for the acousto-persistent current is our
second main result.  It is important to appreciate the difference
between this result and the corresponding one found earlier
\cite{aronov} for the isolated ring. In the isolated ring, the
Holstein process \cite{holstein} required the emission
(absorption) of a specific phonon, with the exact excitation
energy of the electron on the ring. In the present case, the
coupling to the leads turns the bound state into a resonance, with
a width $\Gamma_d^0=-\Im \Sigma_{\rm ext}^R$ which vanishes when
the ring is decoupled from the leads. As a result, there is always
some overlap between the tail of the Green function
$G_{dd}^{0R}(\omega)$ and the Fermi distribution $f(\omega)$,
yielding contributions from Holstein-like processes via phonons
with many energies. Indeed, each contribution to $\Delta I_c$
contains the phase $\delta^0(\omega)$, which vanishes with
$\Gamma_d^0$ ($\delta^0 \sim \Gamma_d^0/|\epsilon_d|$ far from the
resonance). In particular, this results in a non-zero $\Delta I_c$
even at zero temperature: In that limit, if
$\epsilon_d<\mu_\ell=\mu_r=0$, then $n_d=1$. Even with no phonons,
$N_q=0$, the square brackets in Eq. (\ref{ICPH}) become
proportional to $\delta^0(\omega+\omega_q)-\delta^0(\omega)$,
reflecting processes which begin by an emission of phonons. None
of this remains for the isolated ring, when $\Gamma_d^0=0$.

To obtain explicit expressions, we now evaluate the frequency
integration appearing in Eq. (\ref{ICPH}). First, since we operate
within the linear response regime, the voltage is not essential to
our effect and we may safely write in Eq. (\ref{ICPH})
%\begin{eqnarray}
$f_{\ell}(\omega )=f_{r}(\omega )\equiv f(\omega)$
%\end{eqnarray}
%where $f(\omega )=(e^{\beta \omega }+1)^{-1}$ is the thermal Fermi
%distribution (with zero chemical potential) in both reservoirs
\cite{comment1}. Furthermore, we take the electronic temperature
to be low compared to all other energies, so that $f(\omega)
\approx \Theta(-\omega)$. Second, we note that the self-energy
$\Sigma^{R}_{\rm ext}$ takes a simple form when one invokes the
wide-band approximation, with a self-energy which is effectively
at the middle of the (symmetric) conduction band of the leads
\cite{amnon}.  For simplicity, we present results only for the
special symmetric ABI (Fig. 1), with the same hopping matrix
elements on the left and right branches of the ring, and with the
hopping from the ring onto each lead equal to that along the lead.
In the absence of the lower arm of the ABI, the self-energy is
purely imaginary \cite{glazman}, $\Sigma_{\rm ext}^{R}=-i
\Gamma^{0}_{d}$. Adding the lower branch, the self-energy acquires
a real part as well, and both the real and imaginary parts depend
on the flux. That dependence can be written in terms of $T_{B}$
and $R_{B}$, the transmission and the reflection coefficients of
the reference arm alone
%(when the dot is disconnected from the interferometer).
\cite{amnon},
\begin{eqnarray}
\Re\Sigma^{R}_{\rm
ext}&=&-\Gamma^{0}_{d}\sqrt{T_{B}R_{B}}\cos^{2}(\Phi/2),\nonumber\\
\Im\Sigma^{R}_{\rm ext}&=&-\Gamma^{0}_{d}\bigl
(1-T_{B}\cos^{2}(\Phi/2)\bigr ).
\end{eqnarray}
Third, we take the typical phonon frequency to be  much smaller
than the large band-width in the leads. With these approximations
the frequency-integration in Eq. (\ref{ICPH}) is easily performed,
to yield
\begin{eqnarray}
\Delta I_{c}&=&\frac{\Gamma^{0}_{d}}{2\pi}\sin\Phi\sum_{\bf
q}\Bigl [A^{+}_{\bf q}\Bigl
(F(\omega_{q})+F(-\omega_{q})-2F(0)\Bigr
)\nonumber\\
&+&A^{-}_{\bf q}\Bigl (F(\omega_{q})-F(-\omega_{q})\Bigr )\Bigr ],
\end{eqnarray}
where $F(\omega )$ is given in terms of $\delta^{0}(\omega )$, Eq.
(\ref{tan}),
\begin{eqnarray}
F(\omega )=-\sqrt{T_{B}R_{B}}\delta^{0}(\omega )-T_{B}\ln
|\sin\delta^{0}(\omega )|.
\end{eqnarray}
Note again that the dependence of the acousto-persistent current
on the phonon frequency is fixed by the Friedel phase of the dot
at that frequency.  To leading order in the strength of the
electron-phonon coupling, the magnitude of the first term in
$\Delta I_{c}$ is proportional to $A^{+}_{\bf q}$, and thus grows
linearly with the occupation number of the acoustic modes acting
on the dot, $N_q$. In fact, the acousto-persistent current
contains two types of contributions: the part associated with
$F(0)$, which simply represents the 'trivial' Debye-Waller
renormalization of the current, and the novel frequency-dependent
parts, which reflect the change in the persistent current due to
Holstein-like processes.

In summary, we have derived a general expression for the
steady-state current circulating in the Aharonov-Bohm ring, which
is also valid when there are electronic interactions on the dot.
We used this expression to find the effect of an acoustic (or
electromagnetic) source on the persistent current. In particular,
we found that by controlling the intensity of the acoustic wave in
a certain frequency range, one may tune the magnitude of the
orbital moment. The same calculation can also apply to the
photon-induced persistent current.

This project was carried out in a center of excellence supported
by the Israel Science Foundation. It was also partially supported
by the German-Israeli Foundation (GIF) and by the German Federal
Ministry of Education and Research (BMBF), within the framework of
the German-Israeli Project Cooperation (DIP). We acknowledge
helpful discussions with the late A. G. Aronov and with Y.
Levinson.

%and by the Maurice and Gabriella Goldschleger Center for
%Nanophysics  at the Weizmann Institute of Science

\end{multicols}

\begin{references}
\bibitem{comment2} In the nonequilibrium case, $n_{d}$ is determined
by  both the acoustic intensity and  the relaxation processes. We
have extended the relations given in Refs.
\onlinecite{jauho,glazman} to include $n_{d}\ne 0$.
\bibitem{comment1} However, at
thermal electronic equilibrium, $\mu_{\ell}=\mu_{r}$, the
wide-band approximation is not required for the derivation of
$I_{c}$, and the result (\ref{IC}) is exact.
\end{references}
\end{document}